\documentclass[nofootinbib,aps,prb,showpacs,twocolumn,groupedaddress]{revtex4-1}
\usepackage{amsmath}
\usepackage{cancel}
\usepackage{float}
\usepackage{ifsym}
\usepackage{color}
\usepackage{latexsym}
\usepackage{latexsym}
\usepackage{dcolumn}
\usepackage{epsfig}
\usepackage{amsfonts}
\usepackage{amsmath}
\usepackage{dsfont}
\usepackage{accents}
\usepackage[all,cmtip]{xy}
\usepackage{paralist}
\usepackage{enumerate}
\usepackage{amssymb}

\usepackage{bm}
\usepackage{graphicx}
\usepackage{wasysym}
\usepackage{hyperref} % als LETZTES paket laden!!!

\begin{document}

\title{Segmental front line dynamics of randomly pinned ferroelastic domain walls}
\author{S.~Puchberger$^1$}\email{sabine.puchberger@univie.ac.at} \author{V.~Soprunyuk$^1$}  \author{W~Schranz$^1$}
\email{wilfried.schranz@univie.ac.at}
\affiliation{$^1$University of Vienna, Faculty of Physics, Boltzmanngasse 5, A-1090 Wien, Austria} 
\author{M. A. Carpenter}
\affiliation{University of Cambridge, Department of Earth Sciences, Downing Street CB2 3EQ, Cambridge, United Kingdom.}

\date{\today}

\begin{abstract}
Dynamic Mechanical Analysis (DMA) measurements as a function of temperature, frequency and dynamic force amplitude 
are used to perform a detailed study of the domain wall motion in LaAlO$_3$. 
In previous DMA measurements Harrison, \textit{et al}. [PRB \textbf{69}, 144101 (2004)] found evidence for dynamic phase transitions of ferroelastic domain walls in LaAlO$_3$. In the present work we focus on the creep-to-relaxation region of domain wall motion using two complementary methods. We determine, additionally to dynamic susceptibility data, waiting time distributions of strain jerks during slowly increasing stress. These strain jerks, which result from self-similar avalanches close to the depinning threshold, follow a power-law behaviour with an energy exponent $\varepsilon=1.7 \pm 0.1$. Also, the distribution of waiting times between events follows a power-law $N(t_w) \propto t_w^{-(n+1)}$ with an exponent $n = 0.9$, which transforms to a power-law of susceptibility $S(\omega) \propto \omega^{-n}$. The present dynamic susceptibility data can be well fitted with a power law, with the same exponent (n=0.9) up to a characteristic frequency $\omega \approx \omega^{\ast}$, where a crossover from stochastic DW motion to the pinned regime is well described using the scaling function of A.A. Fedorenko, \textit{et al}. [PRB \textbf{70}, 224104 (2004)].      
\end{abstract}

%\keywords{domain wall motion, pinning-depinning transition, avalanches} 

\maketitle

\section{Introduction}
Understanding domain wall motion in ferroic materials is not only of pure scientific interest, but is also important for technical applications\cite{Whyte2015,Gilly2015, Catalan2012, Whyte2014}. Movements of domain walls (DW's) subject to external forces were shown to cause anomalously high values of susceptibility in some ferroelectrics\cite{Mueller2002, Mueller2002-2, Huang1997, Park2000} and ferroelastics\cite{Schranz2003, Schranz2009, Harrison2002, Harrison2003, Puchberger2016} below the phase transition temperature T$_c$. The domain wall response is very sensitive to changes of external conditions, i.e. temperature, frequency, applied field, etc. In some systems, freezing of domain wall motion occurs at temperatures $T_f < T_c $ where the DW's can no longer follow the dynamically applied external force. As a result, the susceptibility drops down to the domain-averaged value. Such behaviour was found for example in dielectric measurements of KH$_2$PO$_4$ (KDP)\cite{Bornarel1992} and (NH$_2$CH$_2$COOH)$_3$ $\cdot$ H$_2$SO$_4$ (TGS)\cite{Huang1996} and in elastic measurements of KMnF$_3$\cite{Kityk2000, Schranz2003}, PbZrO$_3$\cite{Puchberger2016} and LaAlO$_3$\cite{Harrison2002}.\par

As noticed, domain freezing dynamics shares some similarities to glass freezing dynamics \cite{Huang1997}. For example it was found that the relaxation time for domain wall motion follows Vogel-Fulcher behaviour for KDP (T$_{VF} \approx 69~K$), DKDP (T$_{VF} \approx 191~K$) and TGS (T$_{VF} \approx 32~K$). Vogel-Fulcher type domain freezing was also found in KMnF$_3$ doped with 0.003 \% Ca (T$_{VF} \approx 55~K$) \cite{Schranz2009}, whereas Arrhenius behaviour was detected for pure KMnF$_3$ \cite{Salje-Zhang2009}. Meanwhile, Ren, \textit{et al}. \cite{Sarkar2005,Ren2010,Ren2014} found evidence for \textit{strain glass} behaviour in ferroelastic martensites, i.e. Ti$_{50-x}$Ni$_{50+x}$, through a Vogel-Fulcher type relaxation time dependence, typical field-cooling/zero-field-cooling signatures\cite{Ren2010} as well as the observation of dynamic nanodomains which freeze out below $T_g$ at a size of about 20-25~nm. In all these systems, impurities and/or defects seem to play a major role for the freezing process.\par

Very recently, Salje, \textit{et al}. \cite{Salje2014} argued that domain boundary patterns can evolve glass-like states even without any defect induced disorder, which led them to the notion of \textit{domain glass}.
Indeed, large-scale molecular dynamics simulations\cite{Salje2011, Ding2013} of a ferroelastic crystal, with domain walls  
mimicked by a simple two dimensional spring model with a sheared (ferroelastic) ground state, show that DW movements under applied shear deformation follow Vogel-Fulcher behaviour at a certain temperature-regime. They found that pinning/depinning processes also appear as a consequence of domain jamming even if no extrinsic defects are present.\par

Regardless of whether or not defects are present in a sample, there is general consensus that domain wall pinning is a prerequisite for domain freezing. In the domain glass, the twin patterns involve a very high number of twin intersections which act as pinning centres. In other cases, domain walls are pinned at randomly distributed defects. In LaAlO$_3$ the determined values of activation energy suggest that domain walls are predominantly pinned by oxygen vacancies \cite{Harrison2002}.
The basic idea to explain a finite Vogel-Fulcher temperature\cite{Huang1997} $T_{VF}>0$  is then that the pinning becomes correlative with decreasing temperature, leading to an increase of the effective pinning region\cite{Nattermann2004} $\Delta R$. This would imply that the collective pinning energy, $U_{CP}$, diverges at $T_{VF}$ as $U_{CP}=U/(T-T_{VF})$. This is very appealing, since the concept of increasing (with decreasing T) cooperative length scales \cite{Gibbs1965,Karmakar2009}, which leads to a diverging relaxation time at finite temperature, turned out to be very fruitful for glass forming liquids. It is only natural to check if a similar scenario applies also for the domain freezing problem.\par

The main purpose of the present work was to study the pinning-depinning process of ferroelastic domain walls (DW's) in detail as a function of temperature, frequency and applied external force. As an example we used the perovskite crystal LaAlO$_3$, since many aspects of DW movement have been studied\cite{Harrison2002,Harrison2003,Harrison2004,Harrison2010,Harrison2011} and can be used for comparison.\par   

Domain wall pinning effects where already studied some time ago by measuring jerky responses of a system to slowly changing external conditions. For example in ferromagnetic DW's\cite{Durin2006, Colaiori2008, Durin2016} it is known as \textit{Barkhausen noise}. There were serious doubts if a similar crackling noise behaviour could be detected in a crystal with ferroelastic domains, since, due to elastic compatibility, ferroelastic domain walls are rather flat, implying a huge Larkin length and no pinning-depinning transition. 
However, Salje and Harrison \cite{Harrison2010,Harrison2011} found that jerky avalanches also occur during ferroelastic DW propagation. For LaAlO$_3$, the pinning-depinning process was shown\cite{Harrison2011} to be mainly effective at the front line of the needle tips (Fig.\ref{fig:laimag}) which, opposed to the planar parts of the ferroelastic DW's, can easily break into smaller (nanoscale) segments of various length. Recently Gao \textit{et al.}\cite{Gao2014} studied the switching dynamics of individual ferroelastic domains in thin Pb(Zr$_{0.2}$Ti$_{0.8}$)O$_3$ films by using in situ TEM. They found ferroelastic switching mainly to occur at the highly active needle tips in ferroelastic domains. These needle tips are shown to be broken into segments of various length of one to few nm's. \par

Harrison \textit{et al.}\cite{Harrison2010} measured the movement of a single needle domain in LaAlO$_3$ under weak external stress at the critical depinning threshold and found discrete jumps of the needle tip of varying amplitude due to the pinning/depinning of wall-segments to defects. Tracking the movement of the needle tip $x(t)$ yields the dissipated energy via the kinetic energy $E \sim v^2= (dx/dt)^2$. They found that the distribution of energies follows a power law $N(v^2) \propto (v^2)^{-\epsilon}$ behavior with an energy exponent of $\epsilon = 1.8 \pm 0.2$. A similar phenomenon of the jerky movement of \textit{many} DW's in LaAlO$_3$ and PbZrO$_3$ was found recently \cite{Puchberger2017} to have a power law distribution of the maximum drop velocities squared $N(v^2_m) \propto (v_m^2)^{-1.6\pm0.2}$. \par

In the present work we determine additionally the distribution $N(t_w)$ of waiting times between successive jerks, which are related to the energy landscape of the DW segments in the presence of defects (most probably oxygen vacancies in the case of LaAlO$_3$), and compare the calculated complex susceptibilities with frequency dependent elastic susceptibility data. We show that the DW response of LaAlO$_3$ at low frequency of the external stress shows up in three regimes of the complex elastic susceptibility, separated by dynamic phase transitions: sliding at $\omega \tau_{DW} < 1$, stochastic or creep regime (at $\omega \gtrless \omega^{\ast}$) and the pinned regime at $\omega > \omega^{\ast}$. \par     

Section \ref{sec:Experimental} presents details about the samples and the experimental measurement technique. In section \ref{sec:Results} we show temperature dependent elastic susceptibility data at various frequencies, as well as the results of static and dynamic stress scans. We also show waiting time distributions determined from strain jerks at slowly increasing stress and compare the (Laplace transformed) results with dynamic susceptibility data obtained from frequency scans at different temperatures. 
%The data will be analyzed and discussed in section \ref{sec:Discussion}. Section \ref{sec:Conclusion} concludes the paper.  

\section{Experimental} \label{sec:Experimental}

\begin{figure}
\centering
\includegraphics[width=8cm]{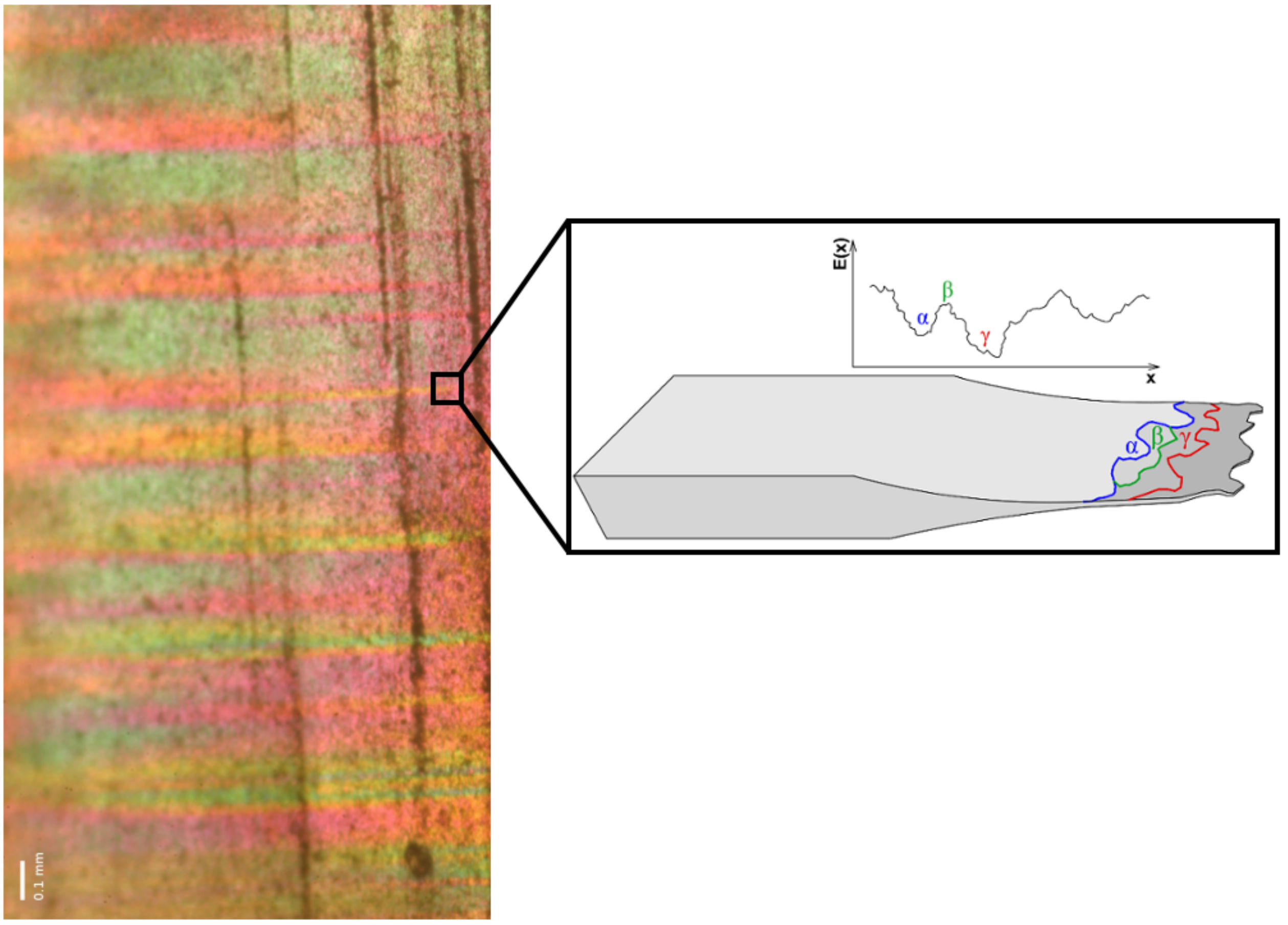}
\caption{Polarizing microscopy image of a LaAlO$_3$ sample and a schematic of a tip of a wedge-shaped needle domain with the propagating front line at the right. The front line is wavy and adapts to the various defect fields.}
\label{fig:laimag}
\end{figure}

For our present study, single crystals of lanthanum aluminate were used. LaAlO$_3$ is a perovskite crystal and exhibits a phase transition to an improper ferroelastic phase. At the phase transition temperature, T$_c = 823K$, the crystal structure changes from cubic Pm$\bar{3}$m to rhombohedral R$\bar{3}$c\cite{Harrison2004}. A typical domain structure of a LaAlO$_3$ sample at room temperature in its rhombohedral phase is shown in Fig.\ref{fig:laimag}. 

Experiments were carried out using the technique of Dynamic Mechanical Analysis (DMA). The measurements were performed under two different operating modes of the DMA: static stress scans and dynamic stress scans of varying frequency, temperature and dynamic force. Static stress scans were performed to measure the sample height h(t) as a function of time and external stress. The external force was slowly increased with time at rates of 3-15mN/min. 

\begin{figure}
\centering
\includegraphics[width=8cm]{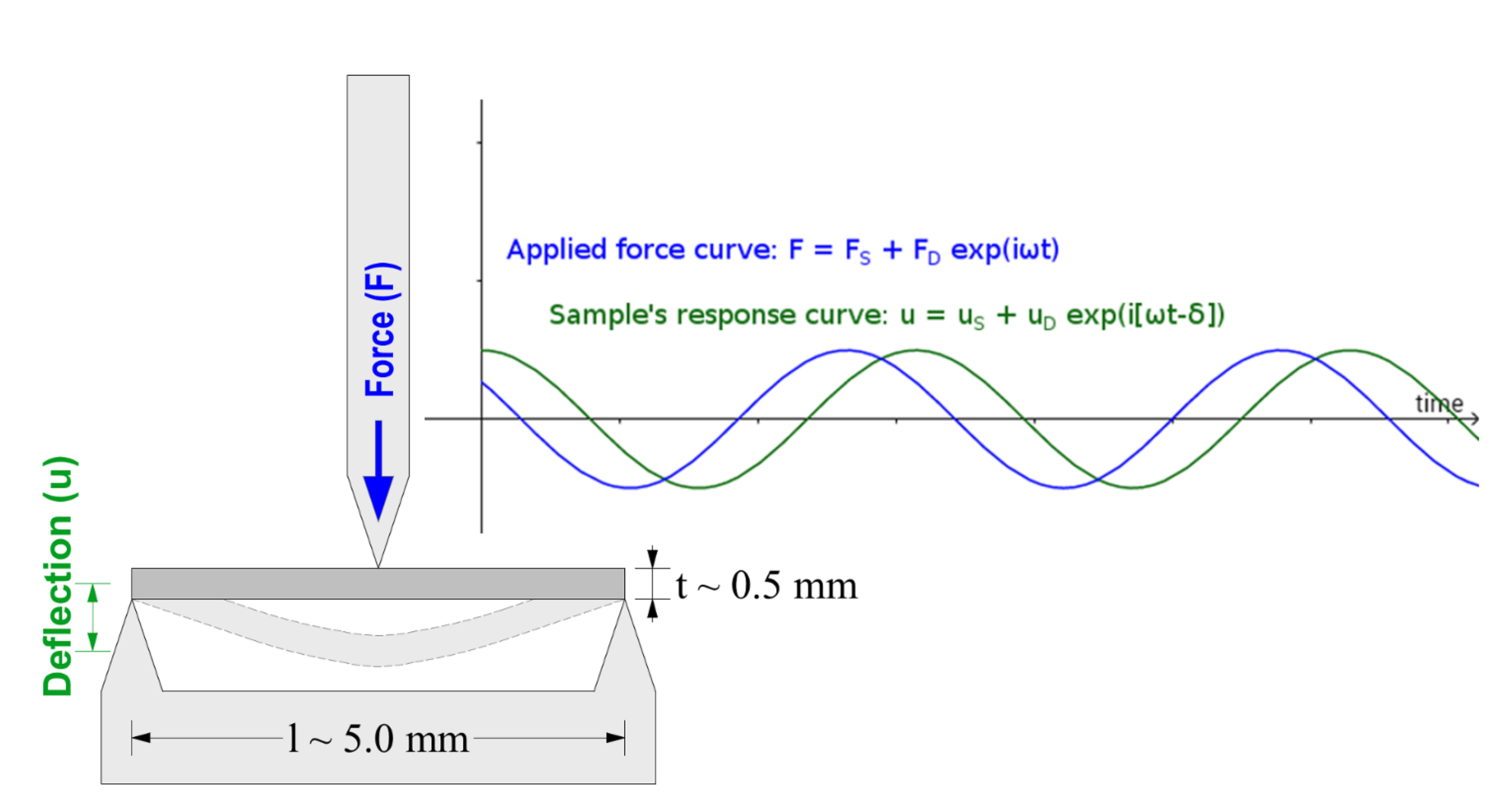}
\caption{Dynamic force and response signals using the DMA technique with dynamic stress mode and three-point-bending geometry of the DMA.}
\label{fig:dmatechnique}
\end{figure} 

By way of contrast, dynamic stress scans involve a sinusoidally varying force, $F_D$. Apart from the dynamic force, a static force, $F_S$ (which is approximately $15\%$  larger than the dynamic force), is applied as well, ensuring that the sample remains in contact with the support edges (see Fig. \ref{fig:dmatechnique}). The DMA measures the amplitude $u$ and phase lag $\delta$ of the mechanical response via electromagnetic inductive coupling and calculates certain components of the real and imaginary parts of the complex elastic compliance, $S^{\ast}$, depending on the orientation of the sample with respect to the applied force. In three-point-bend geometry, the distance between the sample's support edges $l$ is usually much larger than the sample width, $l \gg w$, leading to the complex elastic compliance (in direction $\vec{p}$ perpendicular to the applied force) of the form\cite{Harrison2002}:

\begin{equation}
\label{Eq:comliance}
S^{\ast}(\vec{p})  = \frac{4 t^3 w }{l^3} \frac{F_D}{u} e^{-i\delta}  
\end{equation} 

\noindent 
where $t$ is the sample thickness. 
The complex elastic compliance $S_{ij}$ is related to the elastic constant tensor $C_{ij}$ as $S_{ij} = C_{ij}^{-1}$ and to the real and imaginary parts of the complex Young's modulus $Y^{\ast}=Y'+iY''$ as 

\begin{equation}
\label{Eq:compliance to modulus}
S'=[Y'(1+tan^2\delta)]^{-1} \quad \text{and} \quad S''=Y''| Y^{\ast}|^{-2}
\end{equation}

Static stress scans were conducted using the Pyris Diamond DMA (Perkin Elmer) because this device is able to apply a force up to 10~N, in contrast to the DMA7e which only allows a maximal force of 2.5 N. The resolution of the force is 0.002 N and the resolution of the sample height is about 3 nm. Although the relative accuracy of DMA measurements is about 1\%, the absolute accuracy is usually not better than 20\%. For this reason, all plots are shown here in relative units, i.e. normalized at an appropriate temperature. For dynamic stress scans, the DMA7e (Perkin Elmer) was used because, in contrast to the Diamond DMA, it is possible to set initial values for the dynamic and static forces simultaneously. With the Diamond DMA it is only possible to set an intentional strain. The device regulates static and dynamic stresses according to the sample's stiffness until the required strain is reached.\par

Regarding the sample geometry, three-point-bending was used for all measurements. The LaAlO$_3$ samples were cut in small rods of approximate size 5~x~1.8~x~0.5 mm$^3$, and were placed on two supports with distance 4.2 mm. The force is applied from above halfway along the sample length using an electromechanical force motor.
The maximum temperature used was 620~K, for technical reasons.   

\section{Results} \label{sec:Results}

\subsection{Dynamic stress scans - Dynamic susceptibility}

This section presents the results of dynamic stress measurements in LaAlO$_3$ where both the frequency and dynamic force amplitude were varied. It should be noted that Harrison \textit{et al.}\cite{Harrison2002, Harrison2003} have already performed detailed DMA-measurements on LaAlO$_3$. However, since we intend to compare our strain drop data, i.e. $\epsilon(\sigma(t),T)$, with the DMA data of Y'($\sigma$, f, T) and Y''($\sigma$, f, T), we performed DMA measurements in order to have a complete set of data from the same sample for comparison.\par

\begin{figure}
\centering
\includegraphics[width=8cm]{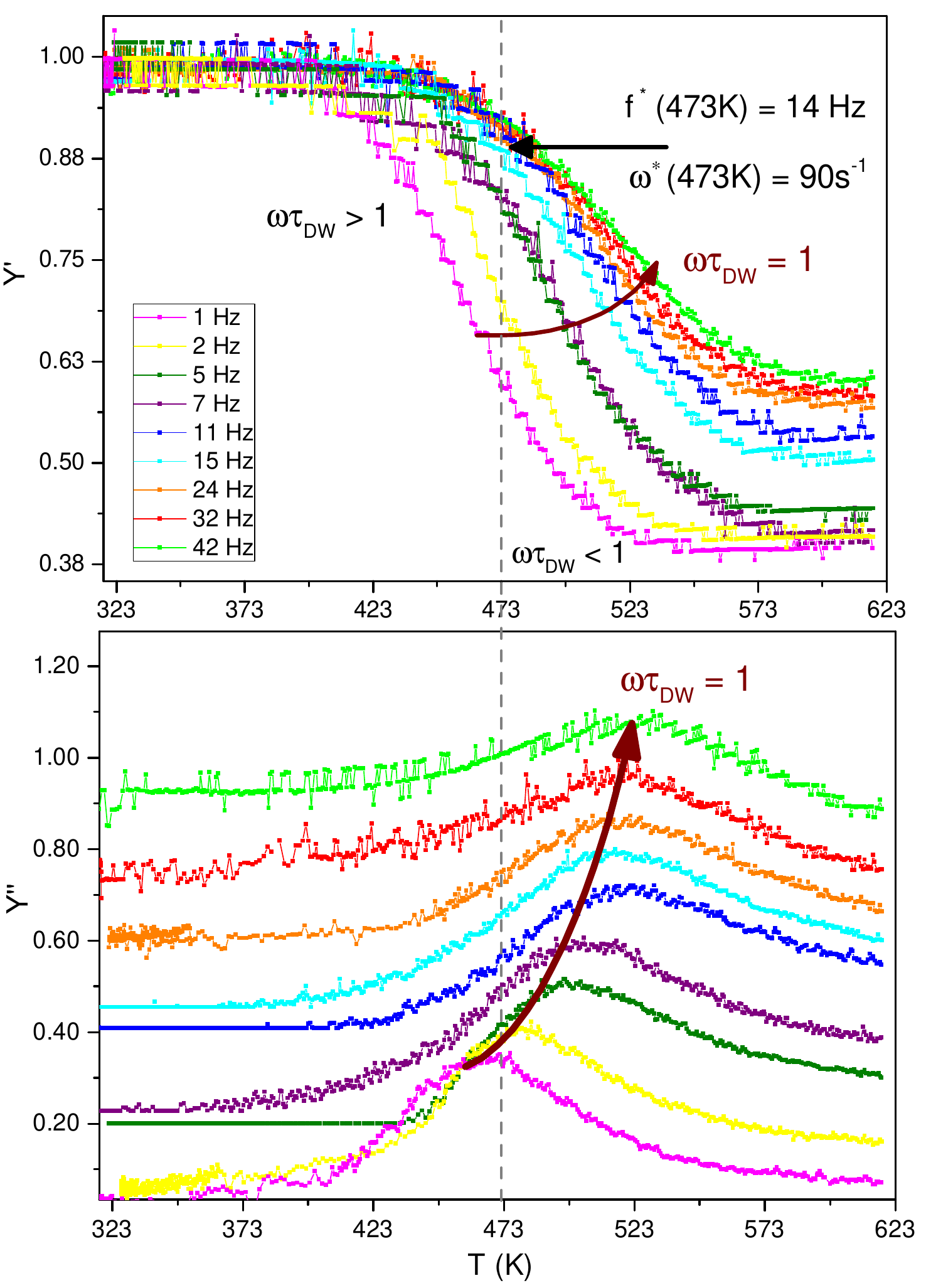}
\caption{Temperature dependencies of the real part of Young's modulus $Y'$ in relative units (top) and the imaginary part $Y''$ (bottom) for LaAlO$_3$ measured at $F_{D}= 300$ mN, $F_{S} = 336$ mN and at different frequencies. $Y''$ curves are shifted for clarity.}
\label{fig:imrefrequ}
\end{figure}

Temperature scans below T$_c$ from room temperature to 623 K with varying measuring frequency are depicted in Fig.\ref{fig:imrefrequ}. During these experiments the dynamic and static stresses were fixed at values of $F_D = 300$ mN and $F_S = 336$ mN. The frequency was changed after each temperature scan. As previously found\cite{Harrison2004}, the low frequency response of the sample at temperatures above $\approx$ 470~K is dominated by domain wall motion in the domain sliding mode ($\omega \tau_{DW}<1$) which induces superelastic softening. At lower temperatures, the DW's gradually freeze out as reflected in an increase of modulus $Y'$ and a peak in $Y''$ at $\omega \tau_{DW}=1$. 
The motion of DW's shows a strong frequency dependence. They can respond to the externally applied stress as long as the characteristic relaxation time $\tau_{DW}$ for DW movement is small enough in comparison with the measurement frequency. With decreasing temperature, $\tau_{DW}$ increases and the DW's can no longer follow the applied stress. If $\omega \tau_{DW} > 1$, this freezing of DW motion is accompanied by a re-hardening of the sample and the elastic response turns to the domain averaged value. The $Y''$- peak shifts to higher temperatures with increasing frequency. For a Cole-Cole relaxation process, the domain wall relaxation time can be extracted from the $Y''$ diagram via determining the shift of the peak maximum, which appears at $\omega \tau_{DW} = 1$. 

\begin{figure}
\centering
\includegraphics[width=8cm]{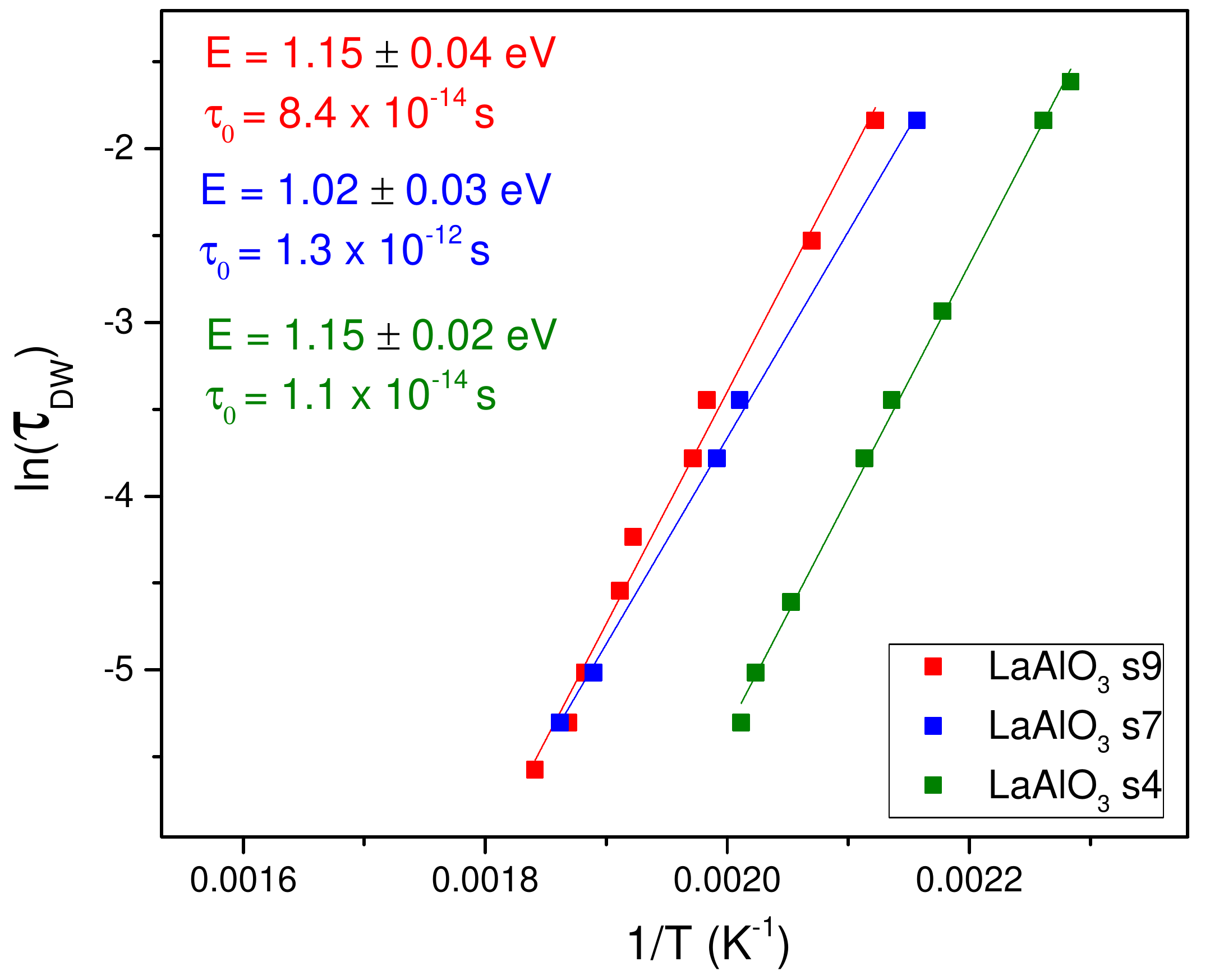}
\caption{Temperature dependence of the relaxation time of domain wall motion plotted in semi-logarithmic scale. The LaAlO$_3$ curves correspond to different samples labeled s4,s7,s9. Samples s9 and s7 were cut from the same piece of LaAlO$_3$, whereas sample s4 was cut from a different piece. The solid lines correspond to Arrhenius-fits. Data points of sample s9 are derived from the curves of Fig.\ref{fig:imrefrequ}.}
\label{fig:relaxationtime}
\end{figure}

A Cole-Cole relaxation is used for fitting $Y''=S''\vert Y^{\ast}\vert^2$ in the crossover region, where
$\omega \tau_{\rm DW}{\rm (T)} < {\rm 1} \rightarrow \omega \tau_{\rm DW}{\rm (T)} > {\rm 1}$, with 

\begin{equation}
\label{Eq:Cole-Cole}
S^{\ast}(\omega) = S_{\infty} + \frac{\Delta S^{\rm DW}}{1 + (i \omega \tau_{\rm DW})^{\mu}}.
\end{equation}

Here $S_{\infty}$ denotes the elastic compliance in the high frequency limit, where $\omega \tau_{\rm DW} >> 1$, and $\Delta S^{\rm DW}$ refers to the DW-induced softening. The exponent $\mu$ leads to a broadening (if $\mu < 1$) of the Debye relaxation, which is obtained in the limit $\mu = 1$.  In agreement with the results of Ref.\onlinecite{Harrison2004} a Cole-Cole function fits the data quite well, if one allows for the broadening parameter $\mu$ to vary between ca. 0.5 and 0.7 as a function of temperature. The relaxation time for LaAlO$_3$ is then well fitted (Fig.\ref{fig:relaxationtime}) with an Arrhenius law

\begin{equation}
\tau_{DW} = \tau_0 exp(E/k_B T),
\end{equation}

\noindent yielding an activation energy, $E = 1.15 \pm 0.04$~eV $\approx$ 110~kJ/mol, for the LaAlO$_3$ sample (s9) used for measurements depicted in Fig.\ref{fig:imrefrequ}. For another sample (s7), a slightly lower value, $E = 1.02 \pm 0.03 $~eV $\approx 96$ kJ/mol, was determined. Both values are quite similar to the results of Harrison et al.\cite{Harrison2004} (E = 0.985, 0.881 and 0.891 eV). 

\begin{figure}
\centering
\includegraphics[width=8cm]{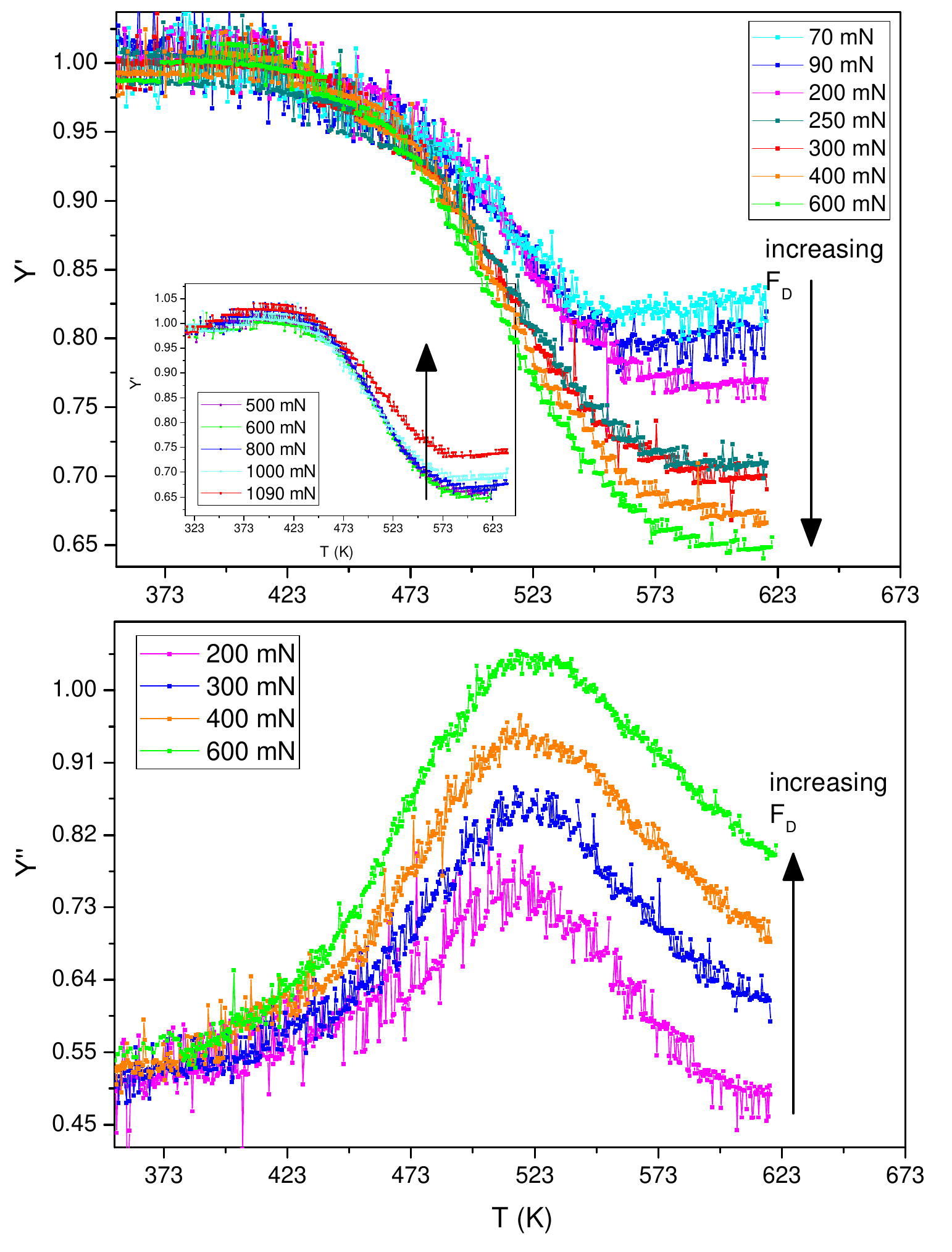}
\caption{Temperature dependencies of $Y'$ of normalized Young's modulus (top) and $Y''$ (bottom) of LaAlO$_3$ measured at 24~Hz and at different dynamical forces ($F_{S} = 1.2 \cdot F_{D}$). $Y''$ curves are shifted for clarity. Inset shows further measurements at increasing forces leading to a re-hardening. }
\label{fig:tscan24hz}
\end{figure}

\begin{figure}
\centering
\includegraphics[width=7cm]{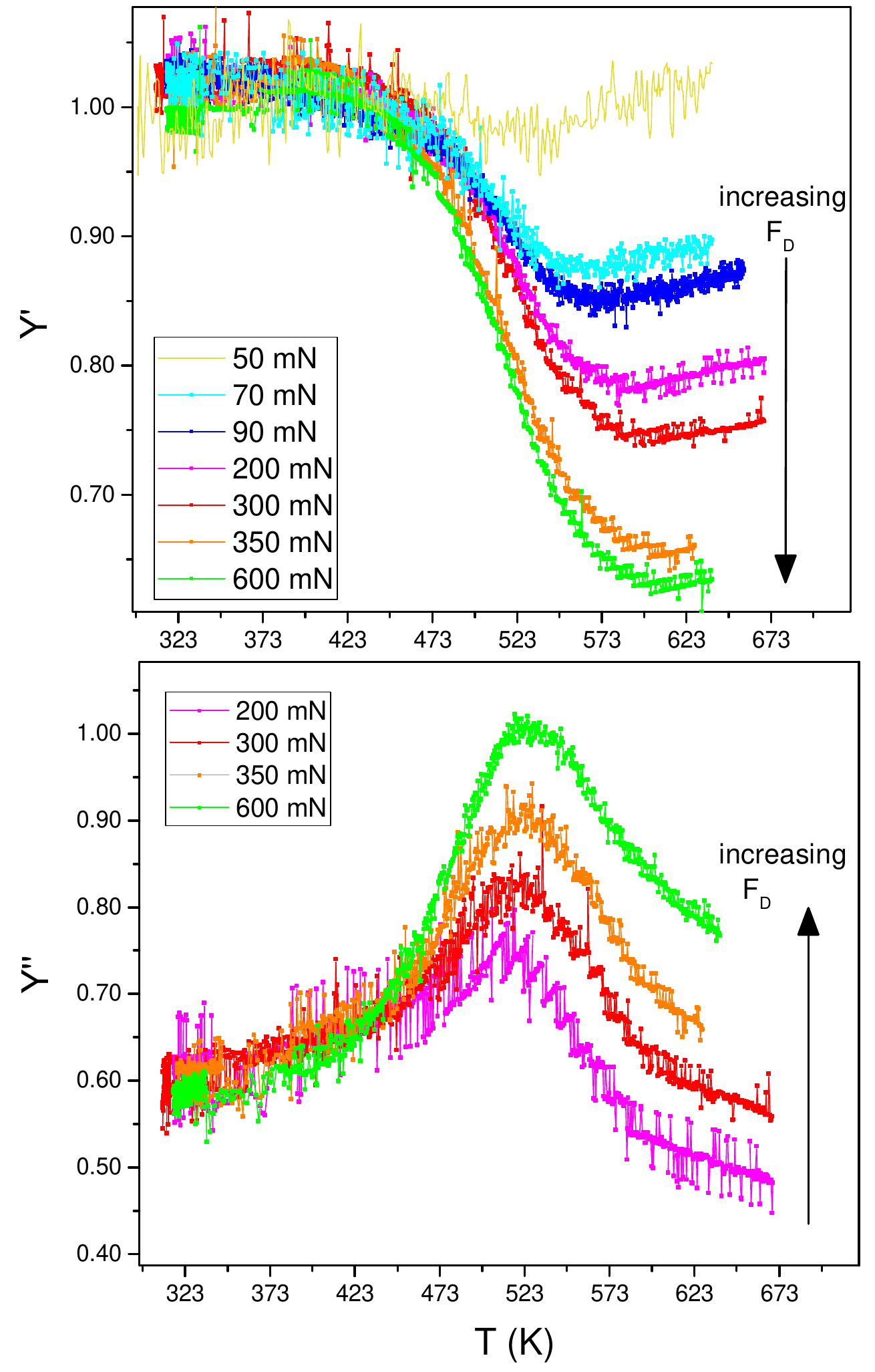}
\caption{Temperature dependencies of $Y'$ of normalized Young's modulus (top) and $Y''$ (bottom) of LaAlO$_3$ measured at 32~Hz and at different dynamical forces ($F_{S} = 1.2 \cdot F_{D}$). $Y''$ curves are shifted for clarity.}
\label{fig:tscan1hz}
\end{figure}

%Fig.\ref{fig:relaxationtime} shows the relaxation time as a function of inverse temperature for the present data of LaAlO$_3$ and for the previously measured PbZrO$_3$\cite{Puchberger2016}. It turns out that $\tau_{DW}$ can be perfectly fitted by an Arrhenius law, whereas clear deviations from Arrhenius are observed for PbZrO$_3$. $\tau_{DW}$ of  PbZrO$_3$ can be better fitted using a Vogel-Fulcher temperature dependence \cite{Puchberger2016}, yet with a much smaller activation energy $E_a \approx$0.23~eV, whose origin is not yet clear.\\

Further investigations of the DW behavior involve variation of the applied external dynamic force. Fig.\ref{fig:tscan24hz} shows results for real $Y'$ and imaginary $Y''$ parts of Young's modulus at different amplitudes of the dynamic force, $F_D$, at a constant frequency of 24 Hz. Increasing the dynamic force amplitude leads to an increasing softening of the sample up to a value of approximately $F_D = 600$ mN. Further increase of the dynamic force amplitude above 600 mN results in a re-hardening (see inset of Fig.\ref{fig:tscan24hz}). Such a behavior is also reflected in an increase of the $Y''$ peak with increasing dynamic force, followed by a decrease at values above 600 mN. A similar pattern is found for other frequencies (see e.g. Fig.\ref{fig:tscan1hz}). The re-hardening at forces $>$600 mN is due to saturation effects, which occur when needle tips retract to the side of the sample where they no longer contribute to the macroscopic strain\cite{Harrison2004}.\par

\begin{figure}
\centering
\includegraphics[width=8cm]{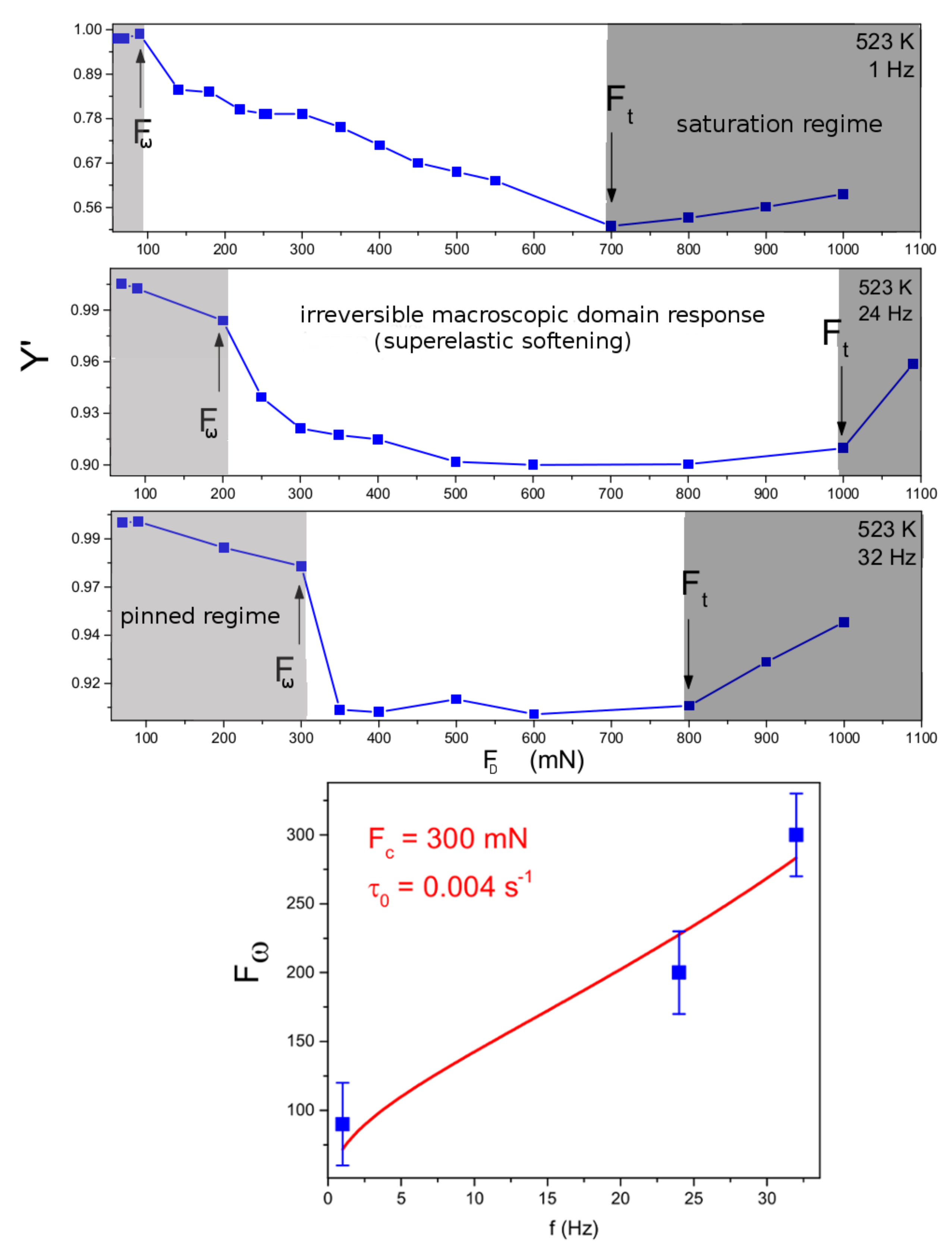}
\caption{Dynamic force dependence of the real part of Young's modulus at 1Hz (top), 24 Hz (middle), 32Hz (bottom) at a temperature of 523~K. The decrease in modulus above a critical value of dynamical force is associated with the unpinning of domain walls, displaying a dynamic phase transition at $F_{\omega}$. The lowest plot shows the critical depinning force $F_{\omega}$ as a function of frequency.}
\label{fig:dynF}
\end{figure}

From the curves in Fig.\ref{fig:tscan24hz} and \ref{fig:tscan1hz}, data points for Fig.\ref{fig:dynF} were extracted to show the real part of Young's modulus as function of dynamic force amplitude for 1~Hz, 24~Hz and 32~Hz at a temperature of 523~K. These plots show that the Young's modulus  decreases rather abruptly at a certain stress value associated with the critical depinning force $F_{\omega}$ which is necessary to set the DW's in motion. Below $F_{\omega}$ the DW's remain pinned. The critical depinning force increases with increasing frequency from about 100~mN at 1~Hz to 200~mN at 24~Hz and 300~mN at 32~Hz.  At forces $F_D > F_{\omega}$, the DW's are able to escape their pinning sites and the superelastic regime is entered. Upon increasing the dynamic force further, $Y'$ decreases and remains at a low value until the dynamic force amplitude exceeds the upper threshold stress, $F_t$, of about 700~mN at 1~Hz and 800~mN at 24~Hz. At stresses above the upper threshold stress, $F_D > F_t$, the saturation regime is reached and the modulus increases again. Hence, the threshold stress separates the superelastic from the saturation regime. Harrison et al.\cite{Harrison2002} showed that the critical depinning stress, $F_{\omega}$, is a function of temperature because thermal fluctuations enable DW's to unpin. The results of the present study demonstrate that $F_{\omega}$ is a function of frequency as well.\par

\begin{figure}
\centering
\includegraphics[width=9cm]{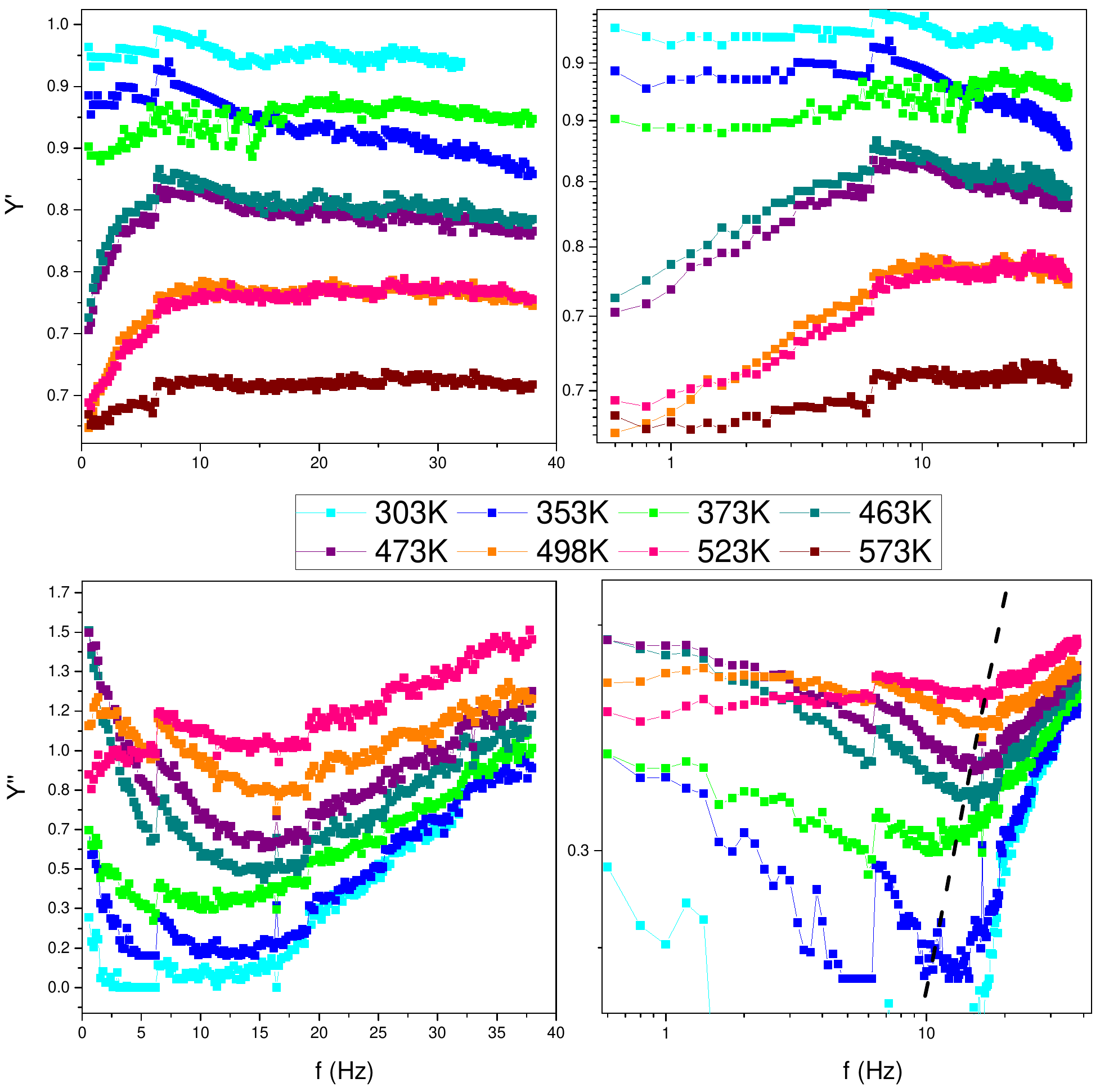}
\caption{Real part $Y'$ of Young's modulus (top) and imaginary part $Y''$ (bottom) as functions of frequency on linear (left) and log-log (right) scale. $F_{S}$ = 448 mN, $F_{D}$=400 mN.}
\label{fig:fscan}
\end{figure}

In addition, frequency scans at constant temperatures, shown in Fig.\ref{fig:fscan}, were performed to further investigate the changes in Young's modulus which occur close to the domain freezing temperature. The static and dynamic forces were fixed at $F_{S}$ = 448 mN and $F_{D}$= 400 mN, and measurements were performed within a temperature range 303~K - 573~K, i.e. starting in the domain freezing regime up to the superelastic regime. At lower temperatures, 303~K - 373~K, the modulus shows hardly any variation with frequency. With increasing temperature, the overall value of the modulus decreases and shows a strong variation of frequency. The dispersion is maximal at temperatures of about 470~K. Increasing the temperature further, the dispersion disappears again.

\subsection{Static stress scans - Strain intermittency}

\begin{figure}
\centering
\includegraphics[width=8cm]{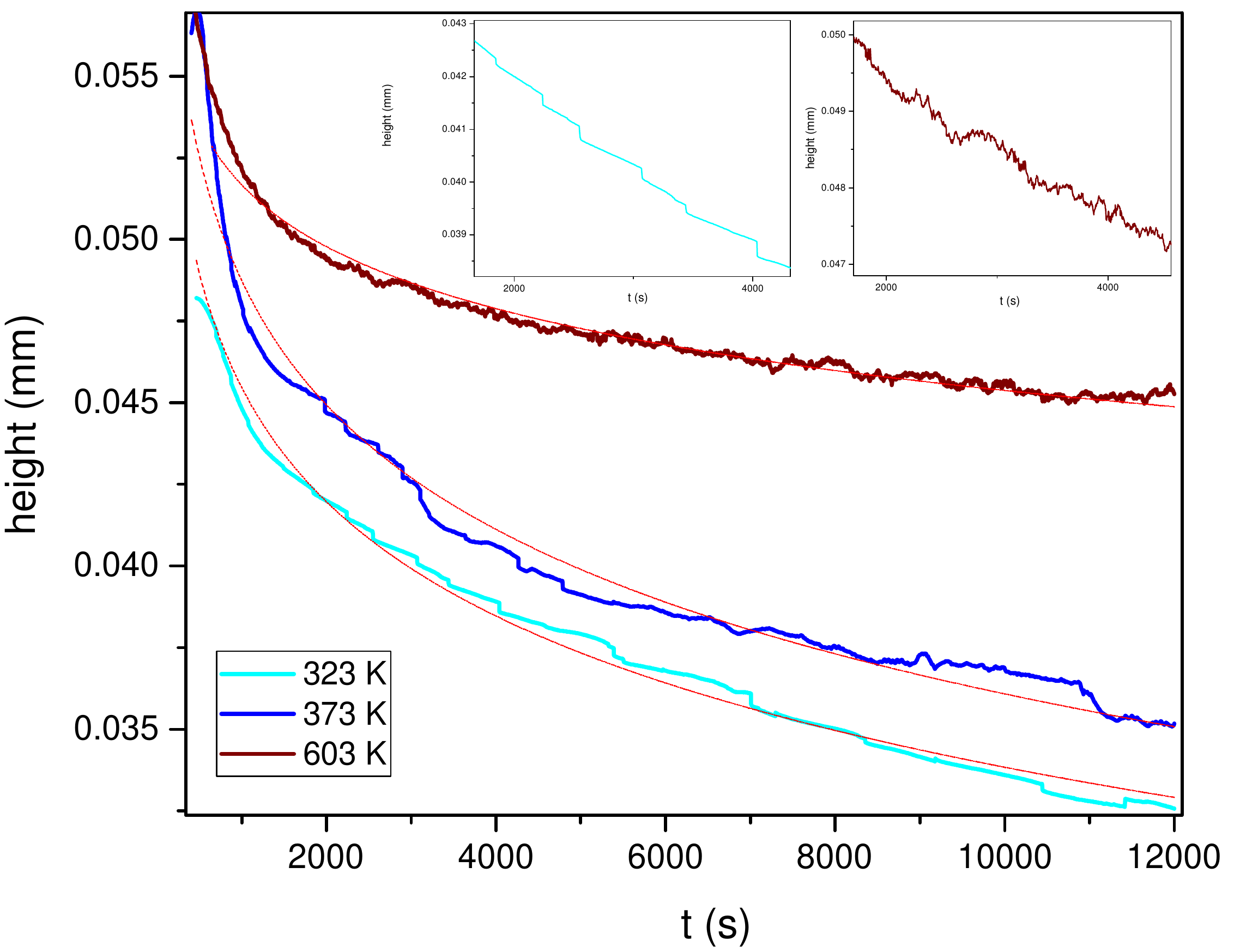}
\caption{Height evolution during a compression experiment of LaAlO$_3$ at different temperatures. The applied force is increased at the rate 10 mN/min from 10 to 2000 mN. (Curves are shifted for clarity.) The red curves correspond to strechted-exponential fits ($\propto e^{-(t/\tau)^\beta}$ with $\beta \approx 0.003$). Insets show magnifications of $h(t)$ revealing different evolutions of height at 323~K (left) and 603~K (right).}
\label{fig:height}
\end{figure}

To study the pinning-depinning process of DW segments in more detail, we have performed static stress scans at various temperatures. Fig.\ref{fig:height} shows the height evolution of LaAlO$_3$ with slowly increasing static stress at different temperatures. At low temperatures (blue curves), the sample height follows a stretched-exponential relaxation envelope punctuated by jerks of varying amplitude. The jerks are manifestations of pinning-depinning events of DW's to defects or due to mutual jamming of DW's. \par

\begin{figure}
\centering
\includegraphics[width=9cm]{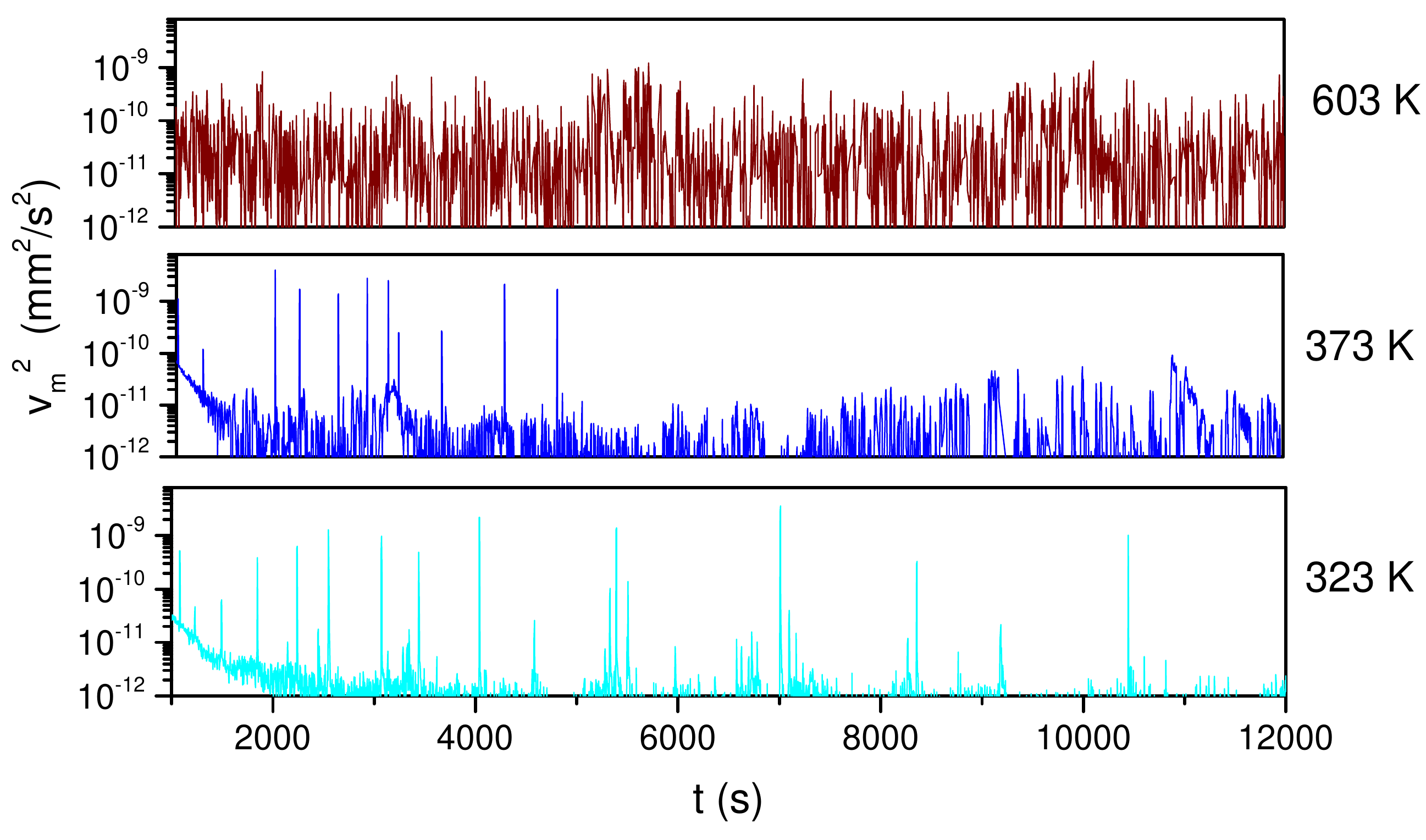}
\caption{Squared drop velocity peaks $v_m^2 = (dh/dt)_m^2$ derived from the height measurements (Fig. \ref{fig:height}) of LaAlO$_3$.}
\label{fig:vsqh}
\end{figure}

Fig. \ref{fig:vsqh} shows the corresponding squared drop velocity peaks derived from the height evolution $h(t)$ with time as $v_m^2 = (dh/dt)_m^2$. They vary over several orders of magnitude and result from about 4000 (at 323~K) single discontinuous strain bursts with about 200 positive velocity jumps, i.e. backward movements of the domain walls. These back-jumps were neglected for further calculations because including them yielded the same statistical results. 

\begin{figure}
\centering
\includegraphics[width=7cm]{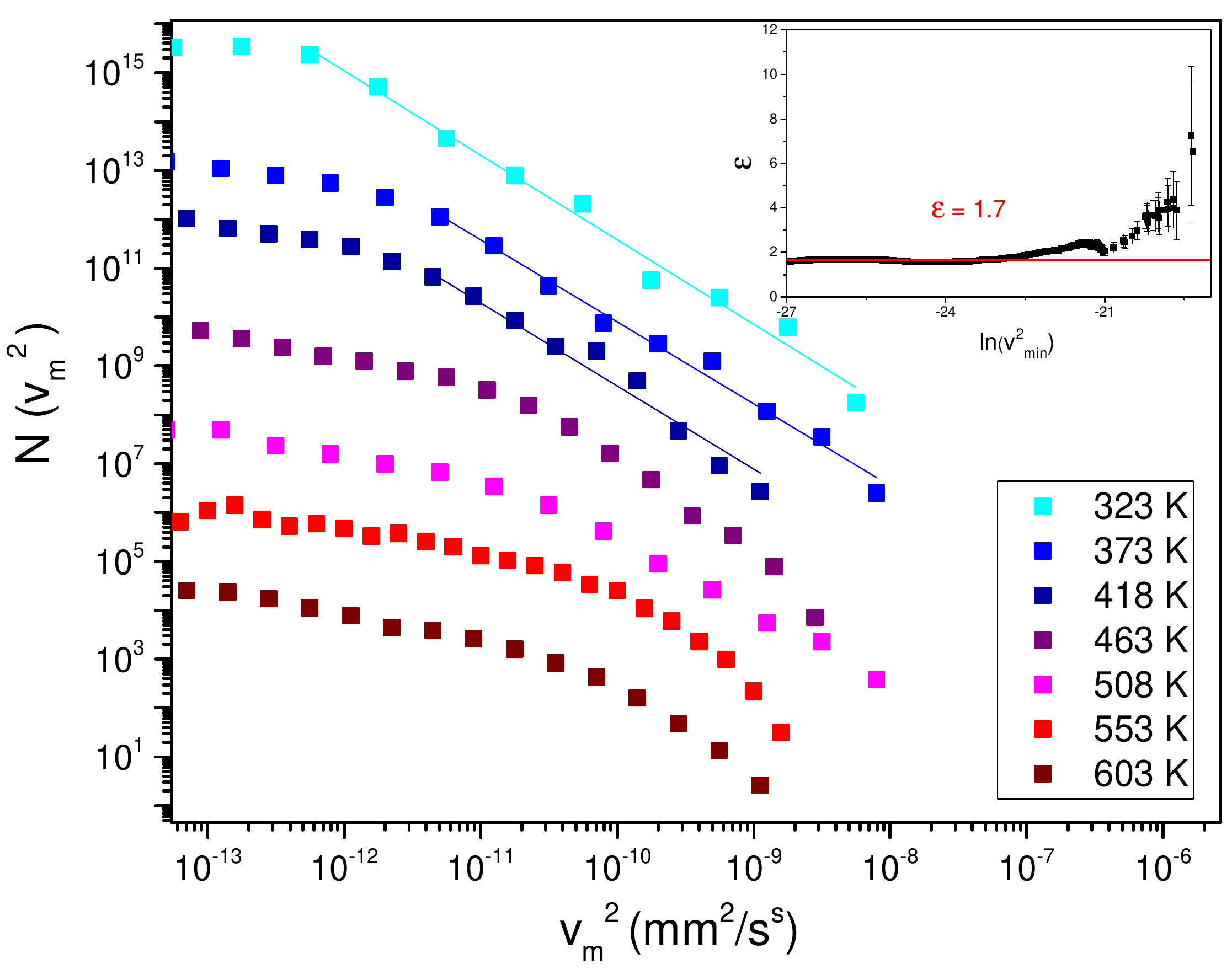}
\caption{Log-log plot of the distribution N(v$^2_m$) of maximum drop velocities squared of LaAlO$_3$ at different temperatures. Curves are calculated from the similar color curves of Fig.\ref{fig:height}. (Curves are shifted for clarity.) The linear fits correspond to power laws with exponent of $\epsilon = 1.7 \pm 0.1$. The inset shows a corresponding maximum likelihood plot.}
\label{fig:Nvsq}
\end{figure}

\begin{figure}
\centering
\includegraphics[width=7cm]{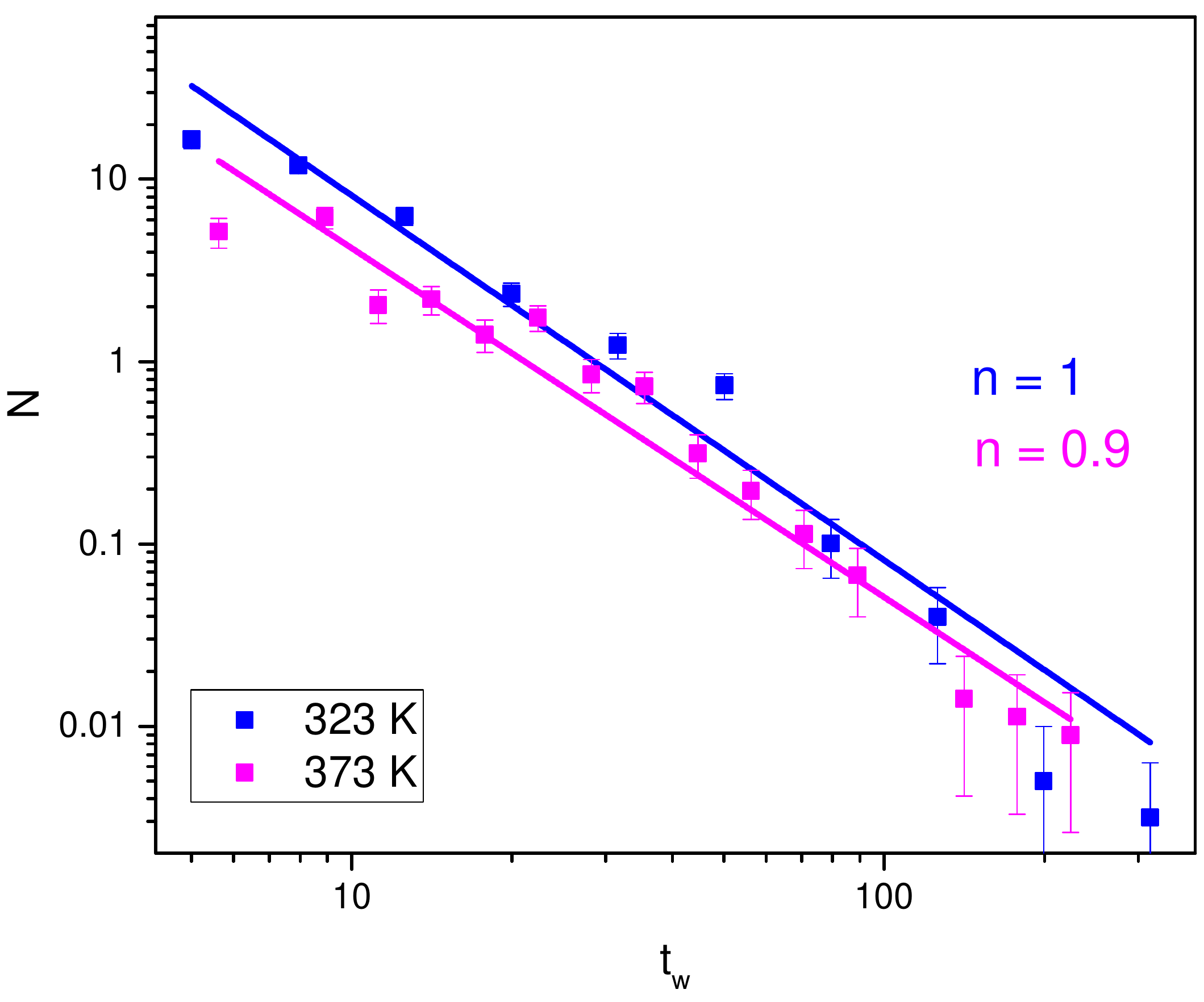}
\caption{Waiting time distributions $N(t_w$) corresponding to measurements shown in Fig.\ref{fig:height} for LaALO$_3$, calculated from the time intervals between successive $v_m^2$ peaks of Fig.\ref{fig:vsqh}.}
\label{fig:waiting}
\end{figure}

For calculation of the power law exponents, the peak data were logarithmically binned (bin size = 0.1) and plotted in a histogram. Fig.\ref{fig:Nvsq} shows the log-log plot of the distributions of squared drop velocity maxima calculated from the statistical characteristics of height drops $\Delta h(t)$. $N(v_m^2)$ is calculated from the squared temporal derivative $v(t)^2 = (dh/dt)^2$ of the sample height $h(t)$. The curves at lower temperatures (curves in different blue shades), i.e. in the frozen regime ($\omega \tau_{DW}>1$) are fitted by a power-law $N(v_m^2) \propto (v_m^2)^{-\epsilon}$ with $\epsilon = 1.7 \pm 0.1$. This exponent value agrees very well with the exponent value of Harrison \textit{et al}.\cite{Harrison2010},  $\epsilon = 1.8 \pm 0.2$, who studied the jerky propagation of \emph{one} needle. \par

At higher temperatures (curves in red shades) the response of the sample differs considerably. An increase in the number of energy jerks (Fig.\ref{fig:vsqh}) with increasing temperature is observed, together with an exponential distribution of $N(v^2_m$). This crossover is in agreement with recent computer simulations of a ferroelastic switching process at different temperatures\cite{Salje2011, Ding2013}, and is most probably due to thermal fluctuations which at high temperature ease the motion of domain wall segments \cite{Nattermann2004} of various length $l_i$ with a rate of $\tau(l_i)^{-1} = \tau_0^{-1} e^{-E(l_i)/T}$.The distribution of waiting times $t_w$ between successive events is shown in Fig.\ref{fig:waiting}. It also yields a power-law $N(t_w) \propto~t_w^{-(n+1)}$, with $n \approx 0.9$.

%Similar results were obtained for PbZrO$_3$. The distributions $N(v_m^2)$ at different temperatures below T$_c$ $\approx$ 503~K are shown in Fig.\ref{fig:Nvsqpzo}. The corresponding squared temporal derivatives $v(t)^2 = (dh/dt)^2$ are depicted together with the calculated waiting time distributions in Fig.\ref{fig:waitingpzo}. 

%\begin{figure}
%\centering
%\includegraphics[width=7cm]{fig13_PZO_Nv.pdf}
%\caption{Log-log plot of the distribution N(v$^2_m$) of maximum drop velocities squared of PbZrO$_3$ at different temperatures. (Curves are shifted for clarity.) The linear fits correspond to power laws with exponent of $\epsilon = 1.6 \pm 0.1$. The inset shows a corresponding maximum likelihood plot.}
%\label{fig:Nvsqpzo}
%\end{figure}

%\begin{figure}
%\centering
%\includegraphics[width=7cm]{fig14_PZO_waitingtime.pdf}
%\caption{Squared drop velocity peaks $v_m^2 = (dh/dt)_m^2$ and waiting time distributions N(t$_w$) calculated from the measurements at 463 K and 373 K of Fig.\ref{fig:Nvsqpzo} for PbZrO$_3$.}
%\label{fig:waitingpzo}
%\end{figure}

\section{Discussion} \label{sec:Discussion}

%The dynamics of elastic interfaces subjected to an alternating force $F_D(t)$ in a random-pinning environment has been a topic of intense research \cite{Vinokur1997,Nattermann1990,Vinokur1996,Feigelman1988,Nattermann2001,Kolton2009}. It is a paradigm for a vast diversity of physical systems, including charge density wave systems\cite{Nattermann1990} driven by an external electric field or cracks in inhomogeneous systems and earth quakes driven by external loading\cite{Baro2013}. 
% Among the first experiments on rough interfaces were those performed on domain walls in disordered magnetic systems \cite{Lemerle1998,Durin2006,Colaiori2008}. Much work has also been done for relaxor ferroelectrics studied by dielectric measurements \cite{Kleemann2003a,Kleemann2003b,Kleemann2005}. An excellent review on the DW dynamics in disordered ferroelectric and ferromagnetic materials was published by Kleemann \cite{Kleemann2007}.

%In the present work we consider the dynamics of domain walls in \emph{ferroelastic} materials LaAlO$_3$ and PbZrO$_3$, showing similar  behaviour as DWs in ferroelectrics and ferromagnetics with varying temperature, frequency and dynamic force amplitude. We show that ferroelastic domain walls are ideal objects for the study of elastic interfaces in random environments.
%In the following we will discuss the present data step by step.

Fig.\ref{fig:imrefrequ} shows real and imaginary parts of Young's modulus of LaAlO$_3$ as a function of temperature at different frequencies. The behaviour is very similar to the DMA-data of Harrison, \textit{et al}. \cite{Harrison2004}. At sufficiently high temperature (say above $\approx 550~K$, depending on $f$) the domain walls can perform macroscopic displacements in response to the applied dynamic force, $F_D > F_{\omega}$, leading to a DW induced superelastic softening at $\omega \tau_{DW} \ll 1$. In this superelastic regime it was shown \cite{Schranz2011} that the DW motion induced elastic compliance $\Delta S^{DW}$ of Eq.(\ref{Eq:Cole-Cole}) can be written as

\begin{equation}
\label{Eq:DWcompliance}
\Delta S^{DW}(T) \propto \frac{N_w}{B}\eta^2(T)
\end{equation}

\begin{figure}
\centering
\includegraphics[width=8cm]{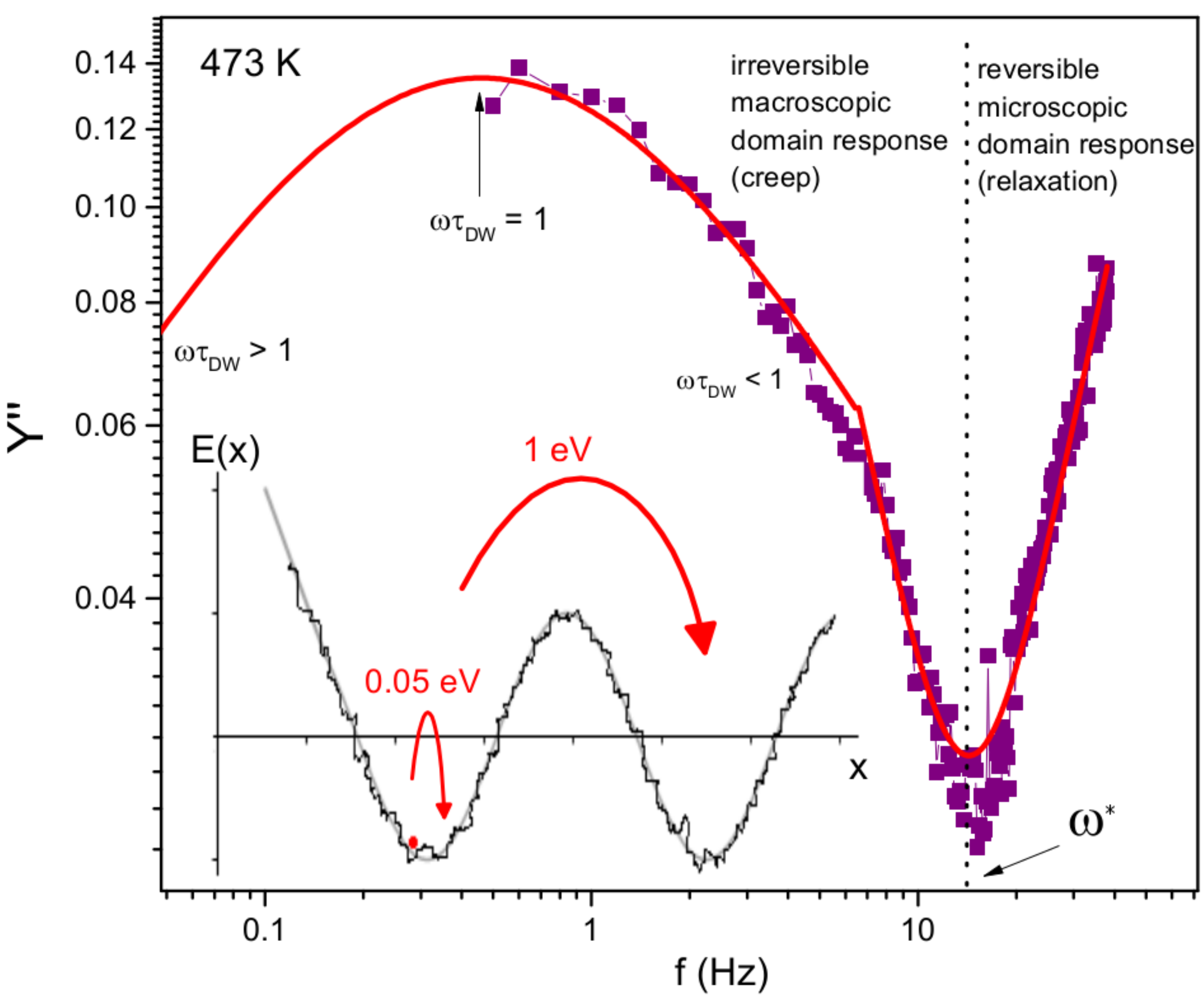}
\caption{Frequency dependence of $Y''$ of LaAlO$_3$ at $T$=473~K, showing two dispersion regions. The part far enough below $\omega^{\ast}/2 \pi \approx$ 15~Hz is well fitted with a Cole-Cole relaxation function. In the region around $f \approx \omega^{\ast}/2 \pi$, a scaling function \cite{Fedorenko2004}, Eq.(\ref{Eq:Federenko}), describes the data quite well.}   
\label{fig:frequencyfit}
\end{figure}   

\noindent where $N_w$ is the number of domain walls, $\eta$ the order parameter and $B$ is the 4th order coefficient of the Landau-expansion. 
Eq.~(\ref{Eq:DWcompliance}) describes the domain wall induced superelastic softening \cite{Schranz2012} in the region $\omega \tau_{DW} \ll 1$ for many ferroelastic systems.\par

In the derivation of Eq.~(\ref{Eq:DWcompliance}) the needle shape of ferroelastic walls plays an essential role. Contrary to ferroelectrics and ferromagnetics, a system of parallel striped ferroelastic domain walls is unstable, because of the lack of a field that corresponds to the depolarization or demagnetization field. However, at the tips of ferroelastic needles, long range elastic fields are created, which act in a very similar way to the stray fields in ferroelectrics or ferromagnetics, and stabilize \cite{Torres1982} an array of ferroelastic needles. The dynamics of such a domain wall array in LaAlO$_3$ have been described phenomenologically\cite{Harrison2004} in the range $\omega \tau_{DW}<1$ by a Cole-Cole function, Eq.~(\ref{Eq:Cole-Cole}). Combining Eqs. (\ref{Eq:Cole-Cole}) and (\ref{Eq:DWcompliance}) we obtain 

\begin{equation}
\label{Eq:DWcompliance with frequency}
\Delta S^{DW} = S^{\infty} + const. \frac{N_w}{B}\eta^2(T) \frac{1}{1+(i\omega \tau_{DW})^{\mu}}
\end{equation}

The behaviour of Fig.\ref{fig:imrefrequ} can be well described with Eq. (\ref{Eq:DWcompliance with frequency}) and $\mu$ varying with temperature between ca. 0.5 - 0.7. Moreover, a fit of the data (e.g. Fig.\ref{fig:imrefrequ}) yields an Arrhenius dependence of the DW relaxation time $\tau_{DW}$ (Fig.\ref{fig:relaxationtime}) with an activation energy $E \approx 1~eV$, which is close to the activation energy commonly associated with oxygen vacancy diffusion in oxide ceramics \cite{Wang2002}. With decreasing temperature, $\tau_{DW}$ increases and a decreasing fraction of DWs can follow the applied dynamic force. The rest of the needles are pinned. This increase of the ratio of static to mobile needle tips with decreasing temperature was observed \cite{Harrison2004} by in situ optical microscopy during DMA measurements.\par

Inspecting Fig.\ref{fig:fscan}, we realize that in the vicinity of the domain freezing regime (around 470~K at 1~Hz), there are at least two dispersion regions. A regime below $f^{\ast} \approx 15~Hz$ and another one above $f^{\ast}$. With the temperature dependence of the domain wall relaxation time $\tau_{DW}$ (Fig.\ref{fig:relaxationtime}), one finds that the region much below $f^{\ast}$ corresponds to the region of macroscopic DW motion. For example at 470~K, $\tau_{DW} \approx  0.06~$s and, accordingly, the region of macroscopic domain wall motion, i.e. where $\omega \tau_{DW} < 1$ to $\omega \tau_{DW} = 1$, is below approximately 3~Hz. Indeed, in the region below ca. 8~Hz, $Y''(f)$ can be well fitted (Fig.\ref{fig:frequencyfit}) with the Cole-Cole relaxation Eq. (\ref{Eq:Cole-Cole}) and the parameters obtained from the fits of the T-dependent measurements (Fig.\ref{fig:imrefrequ}).\par

In the region above $\omega \tau_{DW}=1$, the macroscopic domain wall motion gradually freezes out and only segments of DWs can move in the random potential. For $f<f^{\ast}$ it is assumed that, on the time scale given by $2\pi/\omega$ ($f$ = measurement frequency), the center of mass of the field-driven DW segments probes different local minima of the energy landscape, corresponding to different metastable DW configurations (see inset of Fig.\ref{fig:frequencyfit}). This region has been referred to as the stochastic regime\cite{Fedorenko2004}. We can resort to a large amount of theoretical work\cite{Vinokur1997,Nattermann1990,Vinokur1996,Feigelman1988,Nattermann2001,Kolton2009} to understand the behaviour of DW motion in this region. For example it was shown \cite{Vinokur1997,Fedorenko2004} that the distribution $\psi(t_w)$ of waiting times $t_w(L) = \tau_0 exp[E_B(L)/T]$ for hops of DW segments of length $L$ between metastable states separated by energy barriers $E_B(L) \backsimeq U_c (L/L_c)^{\theta}$ ($U_c$ is the typical barrier on the Larkin scale $L_c$ and $\theta = 2\zeta +D-2$, $D$...dimension of the interface = 2 for DWs, $\zeta$...roughness exponent of DWs = 2/3 for Random Bond impurities\cite{zeta}. This yields $\theta$ = 4/3.) scales as a power law at large times, i.e.

\begin{figure}
\centering
\includegraphics[width=8cm]{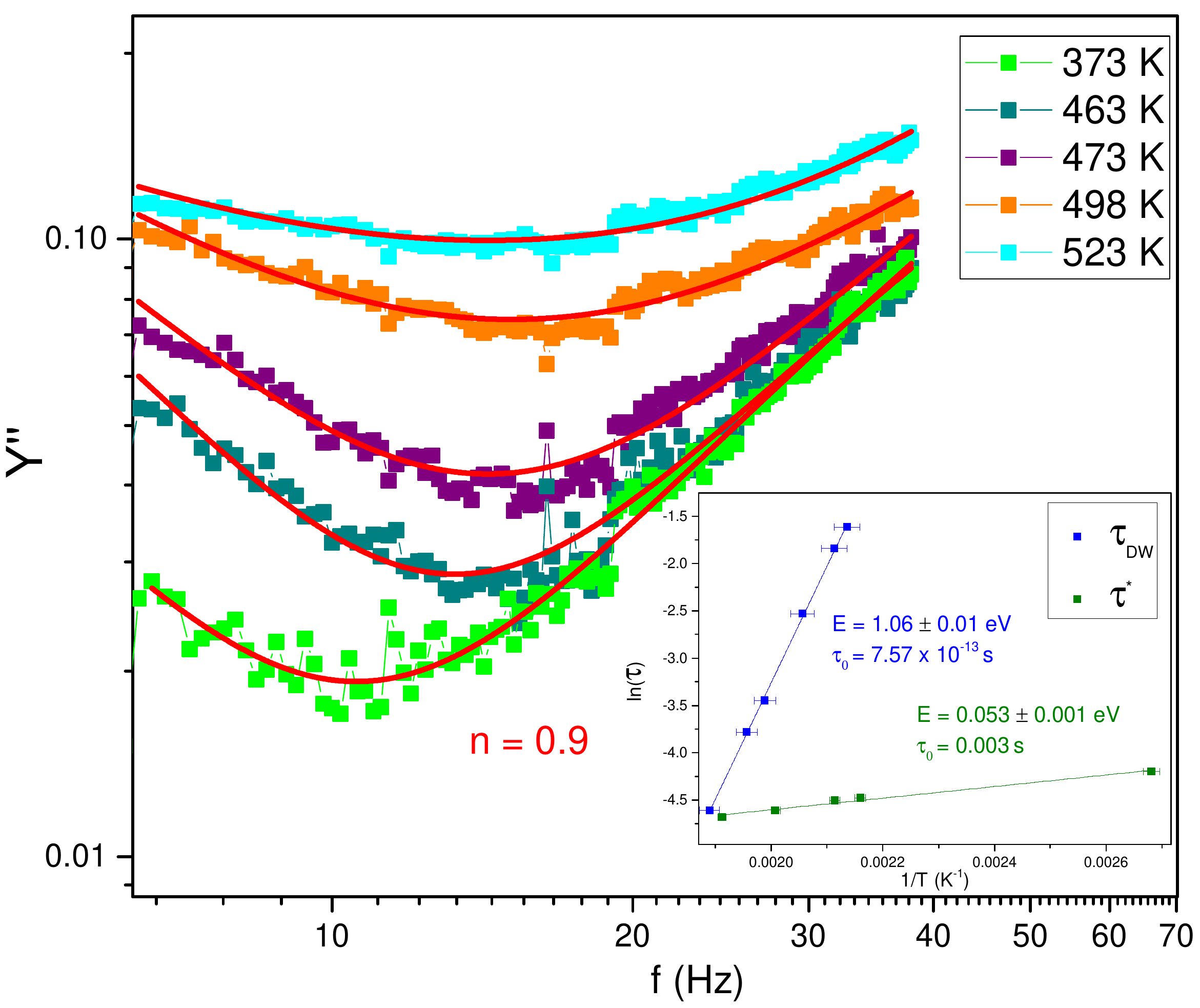}
\caption{Frequency dependence of $Y''$ at various temperatures in the domain freezing regime (data points). The red lines are the best fits using Eq.(\ref{Eq:Federenko}) with $n$=0.9 and $\kappa$=1.2 $\pm$ 0.2. $F_{S}$ = 448 mN, $F_{D}$=400 mN}
\label{fig:Federenkofit}
\end{figure}

\begin{equation}
\label{Eq:distribution of waiting times}
\psi (t_w) \propto (t_w/\tau_0)^{-(n+1)}.
\end{equation}

Here 

\begin{equation}
\label{Eq:power-law exponent}
n=(T/U_c)(\nu -1) ,
\end{equation}

\noindent and $\nu > 2$ determines the size distribution $n(L) \propto L^{-\nu}$ of DW segments. The dynamic response in this regime ($f<f^{\ast}$), which often is called the creep regime, is then given as

\begin{equation}
\label{Eq:creep in frequency domain}
S(\omega) = S^{\infty}[1+(i\omega\tau_0)^{-n}].
\end{equation}

For frequencies corresponding to $\omega \gg \omega^{\ast}$ the DW segments are captured in the valleys, i.e. only relaxational reversible motion of internal modes occurs. In the region $\omega \gtrless \omega^{\ast}$, the imaginary part of the dynamic susceptibility was represented by a scaling function\cite{Fedorenko2004}

\begin{equation}
\label{Eq:Federenko}
S''(\omega) = a\omega^{-n}\left[1+\frac{1}{2\kappa -1}\left( \frac{\omega}{\omega^{\ast}} \right)^{\kappa}   \right]^{2n}
\end{equation}

\noindent Fig.\ref{fig:Federenkofit} shows the measured frequency dependencies of $Y''$ at different temperatures in the domain freezing region. The data can be perfectly fitted using Eq.(\ref{Eq:Federenko}) with $n$=0.9 and $\kappa$= 1.2 $\pm$ 0.2.\par

At this stage it should be stressed that there is perfect agreement between the exponents ($n \approx 0.9$) which are determined by two quite independent experimental methods. The first is from frequency dependent measurements of the dynamic susceptibility $S^{\ast}(\omega)$ at a given temperature (Fig's. \ref{fig:fscan} and \ref{fig:Federenkofit}), and the second is from the intermittent DW response to a slowly increasing stress (Fig's. \ref{fig:height} and \ref{fig:vsqh}), yielding the distribution of waiting times (Fig.\ref{fig:waiting}) between successive jerks. Moreover, using Eq.(\ref{Eq:power-law exponent}) with $n$=0.9, we obtain with $\nu-1 \approx 1$, $T/U_c \approx 0.9$. This implies that, at $T \approx 400~K$, i.e. the temperature, where the crackling noise exponent was measured, the elementary pinning energy is of the order of $U_c \approx 0.04$~eV. Interestingly enough, this value of $U_c$ is rather similar to the value determined from the frequency scans (Fig.\ref{fig:Federenkofit}). 
Indeed, we found an Arrhenius dependence of the depinning frequency $\omega^{\ast} = \tau_0^{-1}exp(-\Delta E/T)$ (see inset of Fig.\ref{fig:Federenkofit}), with an activation energy
$\Delta E \approx 0.05$~eV that is close to the value of $U_c \approx 0.04$~eV. 

%With the threshold frequency \cite{Vinokur1997,Fedorenko2004} $\omega^{\ast}$, separating the sliding regime from the pinned regime     
%\begin{equation} 
%\label{Eq:threshold frequency}
%\omega^{\ast} = \tau_0^{-1} exp\left[-\left(\frac{U_c}{T}\right) \left(\frac{F_c}{F} \right)^{\mu}\right]
%\end{equation}
%we can rationalize this nice agreement, if we identify $\Delta E = U_c (F_c/F)^{\mu}$, assuming $(F_c/F)^{\mu} \approx 1$. Along the same line of arguments 

Along the same line of reasoning, we can also understand the frequency dependence of the threshold force $F_{\omega}$ (Fig.\ref{fig:dynF}), which was shown\cite{Nattermann2001} to obey the relation 

\begin{equation} 
\label{Eq:Fomega}
F_{\omega} = F_c  \left[  \frac{U_c}{T ln(\frac{1}{\omega \tau_0})} \left(1-\frac{F_{\omega}}{F_c}\right)^{\theta}\right]^{1/\mu}.
\end{equation}

With $\theta$ = 4/3 for random bond impurities, and $\mu = \theta/(2-\zeta) = 1$ for $\zeta$ = 2/3, we can approximate Eq.\ref{Eq:Fomega} as

\begin{equation} 
\label{Eq:Fomega2}
F_{\omega} =  \frac{F_c}{1 + \frac{T\cdot ln(1/(\omega \tau_0)}{U_c}},
\end{equation}

\noindent which describes the observed (Fig.\ref{fig:dynF}) increase of $F_{\omega}$ with increasing $\omega$ rather well. Eq. \ref{Eq:Fomega} also explains the $\sim 1/T$ dependence of the depinning force found by Harrison, \textit{et al.} (Fig.10 of Ref.\onlinecite{Harrison2002}). \par

The observed similarity between $\Delta E \approx U_c$ can be well explained with the threshold frequency\cite{Fedorenko2004}

\begin{equation}
\omega^{\ast} = \tau_0^{-1} exp\left [-\frac{U_c}{T}\cdot \left (\frac{F_c}{F}\right )^{\mu}\right ] = \tau_0^{-1} exp \left (\frac{\Delta E}{T} \right )
\end{equation}

\noindent and the value $F_c\approx300$~mN, obtained from the fit (Fig.\ref{fig:dynF}). During frequency scans (Fig.\ref{fig:Federenkofit}) the applied force was $F\approx 400$~mN. The values of $F_c\approx$~300~mN and $F\approx~400$~mN support the assumption $\Delta E~=~\frac{U_c}{T}(\frac{F_c}{F})^{\mu}~\approx~\frac{U_c}{T}$.

%Although, we do not know the values of F$_c$, $\theta$ and $\mu$, one easily finds a range of parameters ($F_{\omega} < F_c$), showing that F$_{\omega}$ increases with increasing frequency of the alternating force (see Fig.\ref{fig:dynF}). 

\section{Conclusions} \label{sec:Conclusion} 
Up to this time, work on "elastic" interfaces in random environments of ferroics has been mainly focused on ferromagnetic and ferroelectric systems. In the present study we investigated the ac-response of elastic DWs in LaAlO$_3$. Similarly to many other systems, where a competition between disorder (due to defects) and order (due to interfacial elasticity) leads to a rugged energy landscape with many metastable states, this is also the case for ferroelastic DWs in the presence of defects. \par

By measuring the complex linear susceptibility of LaAlO$_3$ at low frequency we found clear effects of such a complex energy landscape. The data can be well modeled within a scaling approach by taking account of local pinning and motion of DW segments under random pinning forces\cite{Fedorenko2004, Nattermann1990}.  At temperatures $T \approx T_f$ around the domain freezing regime, where the macroscopic motion of DWs has already stopped during one period of the alternating force ($\omega \tau_0 > 1$), segments of DWs of length $L$ can still overcome local barriers of height $E_B(L)$, even at forces $F_{\omega} < F_c$, where $F_c$ is the pinning force at $T$=0~K. This leads to an irreversible creep like wall motion with $S''(\omega) \propto \omega^{-n}$ with $n\approx$ 0.9. To study this non-Debye response also in the time domain, we measured the distribution of waiting times $\psi(t_w)$ needed to overcome the energy barriers, which, according to theory (e.g. Ref.\cite{Vinokur1997}), should scale as $\psi(t_w)\propto t_w^{-(n+1)}$ . Although these measurements of $\psi(t_w)$ are rather complementary (based on strain bursts during slow compression) to the frequency dependent susceptibility measurements, there is a remarkable agreement between both methods, both lead to $n$=0.9. \par

In summary, the present results suggest that DW dynamics in disordered ferroic materials are rather universal. Moreover, ferroelastic domain walls are ideal objects to study the dynamics of elastic manifolds driven through a random medium. Dynamic mechanical analysis is a very appropriate method for its study, since it can be used in a complementary way to provide both the dynamic elastic response in the frequency domain (dynamic susceptibility) and in the time domain (through strain intermittency measurements). \par

Further measurements on various systems have to be done to understand the DW dynamics around $\omega \tau_{DW} \approx 1$ in more detail, so as to reveal the microscopic origin of domain freezing and to see if the observed deviations of $\tau_{DW}$ from Arrhenius behaviour, found in some systems, are manifestations of a domain glass.

\noindent \textbf{Acknowledgments} The present work was supported by the Austrian Science Fund (FWF) Grant No. P28672-N36.

\end{document}